\documentclass[prl,reprint,showpacs,twocolumn,superscriptaddress,floatfix]{revtex4-1}

\usepackage{epsfig}
\usepackage{amsmath}
\usepackage{amssymb}
\usepackage{amsfonts}
\usepackage{mathptmx}
\usepackage{dcolumn}
\usepackage{eucal}
\usepackage{bm}
\usepackage{color}
\usepackage[colorlinks,linkcolor=blue,citecolor=blue]{hyperref}
\usepackage{epstopdf}
\usepackage{bbold}
\usepackage{url}

\usepackage{dsfont}

\def\>{\right\rangle}
\def\<{\left\langle}
\def\be{\begin{equation}}
\def\ee{\end{equation}}
\def\ba{\begin{array}{lll}}
\def\ea{\end{array}}

\def\beq{\begin{eqnarray}}
\def\eeq{\end{eqnarray}}

\def\d{{\rm d}}
\def\i{{\rm i}}
\def\e{{\rm e}}

\newcommand{\average}[1]{\langle #1 \rangle}
\newcommand{\Trace}{{\rm Tr}}

\newcommand{\D}{\mathcal{D}}

\newcommand{\matrixel}[3]{{\mathinner{\langle{#1}| {#2} | {#3}\rangle}} }

\begin{document}

\title{Energy exchange in driven open quantum systems at strong coupling}

\author{ Matteo Carrega}
\affiliation{SPIN-CNR, Via Dodecaneso 33, 16146 Genova, Italy}
\author{ Paolo Solinas}
\affiliation{SPIN-CNR, Via Dodecaneso 33, 16146 Genova, Italy}
\author{ Maura Sassetti}
\affiliation{Dipartimento di Fisica, Universit\`a di Genova, Via Dodecaneso 33, 16146 Genova, Italy}
\affiliation{SPIN-CNR, Via Dodecaneso 33, 16146 Genova, Italy}
\author{ Ulrich Weiss}
\affiliation{2. Institut f\"ur Theoretische Physik, Universit\"at Stuttgart, D-70550 Stuttgart, Germany}

% 03.65.Yz Decoherence; open systems; quantum statistical methods
% 05.60.Gg Quantum transport

%\email{teo.carrega@gmail.com}

\begin{abstract}
The time-dependent energy transfer in a driven quantum system strongly coupled to a heat bath is studied within an
influence functional approach. Exact formal expressions for the statistics of energy dissipation into the different channels 
are derived. The general method is applied to the driven dissipative two-state system. It is shown that the energy  flows obey a balance relation, and that, for strong coupling, the interaction may constitute the major dissipative channel.
Results in analytic form are presented for the  particular value $K=\frac{1}{2}$ of strong  Ohmic dissipation. The energy flows show interesting behaviors including driving-induced coherences and quantum stochastic resonances. It is found that the general characteristics persists  for $K$ near $\frac{1}{2}$.
\end{abstract}
\maketitle

{\it Introduction ---}
Deep understanding and precise control of energy exchange in strongly  coupled systems at the quantum level may  have profound impact both from a fundamental and a practical point of view. On one side, it can advance the formulation of a  consistent nonequilibrium thermodynamics  for strong coupling. 
In fact, the question how to conceive a consistent thermodynamics for nonequilibrium scenarios with strong system-reservoir coupling has attracted recent interest \cite{campisi2009fluctuation, esposito2015quantum, esposito2015nature, strasberg2016nonequilibrium, talkner2009fluctuation}, and the issue is currently a subject of controversy.
Substantial progress has recently been made for classical systems \cite{seifert2016first}.
On the applicative side, it can have great impact on  the active field of quantum optimal control with pioneering applications in biology and information technology  \cite{Herek2002Quantum,Caruso2012Coherent, oviedo-casado2016Phase}.

The natural setting to study energy exchange and dissipation is to consider an externally driven quantum system coupled to one or several heat reservoirs~\cite{gasparinetti2014fast, viisanen2015, schmidt2015, hekking2013quantum, suomela2014, ankerhold2014}. This might also serve as a building block in the study of more complex systems.
Usually, the system-reservoir entity is studied in the weak-coupling limit. This is technically advantageous, since perturbative methods can be used. The weak coupling assumption is well justified in the theoretical analysis of experiments in which one would like to have long coherence times. 
There are, however, many physical systems for which the interaction energy is comparable with the system energy.
Among all, the most striking ones are e.g. a biomolecule in a solvent and photosynthetic systems, in which energy is absorbed and efficiently transported over long distances. It is now well-established that the vast efficiency of this process cannot be explained with concepts of classical charge transport. Rather it exhibits not only quantum features \cite{Engel2007, Collini2010, Panitchayangkoon2010}  but strong interaction with a structured non-Markovian reservoir seems to be crucial for the high transport efficiency \cite{mohseni2008environment, Lloyd2011,oviedo-casado2016Phase, Guarnieri2016Energy}.

With such motivations, it is natural to ask if similar spectacular effects may arise in the energy transfer characteristics
of small and well-controlled quantum systems, when they are externally driven and strongly coupled to a heat reservoir. 
The time-dependent force is continuously pumping  energy into the system, and thus drives it out of equilibrium~\cite{ludovico2014dynamical, ludovico2016adiabatic}.
Coincidently, the strong coupling with the environment makes the dynamics more  intricate, 
since the system can dissipate energy through different channels.
%Furthermore, the impact of strong quantum noise in a driven quantum system may lead to interesting effects, such as e.g. quantum stochastic resonance  (QSR) phenomena, in which an external signal is amplified because of the noise\cite{grifoni1996coherent, grifoni1998driven}.
The convoluted interplay between all these competitive contributions and the identification of the dissipative
channels involved is the subject of this Letter.

First, we set the stage by facing the problem from a very general perspective. To this end we establish a functional a
functional integral approach for the characteristic function of the energy transfer statistics, in which the dissipative channels to the reservoir and to the interaction are identified.
Then we apply this method to the versatile spin-boson system. We establish the energy balance relation and 
analyze the inherent dissipative channels. Results in analytic form are presented  for the particular value $K=\frac{1}{2}$
 of strong Ohmic 
dissipation \cite{tsvelick1983exact, leggett1987dynamics, sassetti1990, weiss2012quantum}. It is shown that the system-reservoir coupling can be the dominant dissipative channel in particular regions of the parameter space. It is found that the drive-induced coherences and quantum stochastic resonance features of the model \cite{grifoni1996coherent,  grifoni1998driven, kast2013persistence, kast2013dynamics} are reflected in the energy transfer characteristics.
Finally, it is shown that the characteristic features  also hold for $K$  near $\frac{1}{2}$.

{\it General approach to energy exchange ---}
The Hamiltonian of the system-plus-reservoir is
$ H(t) = H_{\rm S}(t)+ H_{\rm R}+ H_{\rm I}$ with the
system part $H_{\rm S}(t) = H_0+V(q,t)$, the reservoir part  $H_{\rm R} = 
\sum_{\alpha}  \Big [ \frac{p_\alpha^2}{2 m_\alpha } + \frac{1}{2} m_\alpha \omega^2_\alpha x_\alpha^2 \Big ]$, 
and the translational-invariant coupling  $H_{\rm I} = - q \sum_{\alpha} c_\alpha 
x_\alpha +\frac{1}{2}\sum_\alpha \frac{c_\alpha^2}{m_\alpha\omega_\alpha^2}$~\cite{caldeira1983path, weiss2012quantum, ingold2002path, leggett1987dynamics}. 
The driving is carried by the time-dependent potential $V(q,t)$.
All effects of the reservoir coupling on the system are captured by the spectral density 
$J(\omega) = \frac{\pi}{2}\sum_\alpha
\frac{c_\alpha^2}{m_\alpha\omega_\alpha^2}\delta(\omega-\omega_\alpha)$.

We are interested in the energy flow of the driven system coupled strongly to the heat bath.
In the usually studied weak coupling limit, it is assumed that all the energy is dissipated into the reservoir \cite{campisi2011colloquium, campisi2011erratum, carrega2015functional, esposito2009Erratum, esposito2009nonequilibrium, schmidt2015, hekking2013quantum}.
In this case, the dissipated energy is found out by a double projective measurement of the reservoir energy at the beginning and at the end of the evolution.

If the weak coupling limit is not met, the system-bath interaction represents an additional dissipative channel. In the
sequel, we present an approach which allows us to analyze the time-dependence of the  various energy transfer  contributions.

In a complete description of the energy transfer, knowledge of both the variations of the  energy dissipated into
the reservoir,
$\langle{\cal E}_{\rm R}(t)\rangle \equiv \langle H_{{\rm R}}(t)\rangle - \langle H_{{\rm R}}(0)\rangle$, and into 
reservoir-plus-interaction,  $\langle{\cal E}_{{\rm RI}} (t)\rangle \equiv \langle H_{{\rm R}}(t)\rangle +\langle H_{{\rm I}}(t) \rangle  - \langle H_{{\rm
R}}(0)\rangle - \langle H_{{\rm I}}(0)\rangle$, is  essential. {Here the average of the observable $O$ is defined as $\langle O(t)\rangle
= {\rm Tr}[O\rho(t)]$,
 where $\rho(t)$ is the density operator.}
 Moreover the variation of energy related to the interaction can be inferred simply by the difference $\langle {\cal E}_{{\rm I}}(t)\rangle \equiv  \langle {\cal E}_{{\rm RI}}(t)\rangle - \langle {\cal E}_{{\rm R}}(t)\rangle$.
It is worth to underline that these observables can be related to specific measurement protocols.
Assume that at time $t\le 0$ the system and the bath are decoupled and the system dwells in a diagonal state
of the reduced density matrix (RDM) with probability distribution $p_{\rm S}^{}(q(0))$. 
Initially we have
$\langle q(0) |\rho(0)|q(0)\rangle= p_{\rm S}^{}(q(0))\, \e^{-\beta H_{\rm R}}/ {\rm Tr}[\,\e^{-\beta H_{\rm R}}]$, where $\beta=1/T$ is the inverse temperature (throughout we put $k_{\rm B}^{}=\hbar=1$).
Immediately before both the coupling with the reservoir and driving force are switched on, at time $t=0$, the bath energy is measured.
At a later time $t$, we either switch off the coupling and measure the bath energy again, or we coevally measure the 
energy of the bath and the coupling. 
These measurements may be projective \cite{gasparinetti2014heat, silaev2014lindblad, carrega2015functional} or performed with a full quantum detector \cite{solinas2015fulldistribution}. The moment generating function (MGF) $G_\lambda (\nu , t)$ embodies the entire statistics of these measurement protocols. Importantly, it can be written as the trace of a generalized density operator~\cite{esposito2009nonequilibrium, gasparinetti2014heat},
\begin{equation} \label{eq:gf1}
  G_\lambda (\nu,t) = \Trace\, [ \e^{\i \nu( H_{\rm R}+\lambda H_{\rm I})} U_{t, 0} \,
\e^{- \i  \nu  H_{\rm R} }\rho(0) U^\dagger _{t, 0} ] \, .
\end{equation}
Here the operator $U_{t,0}$ implements the {unitary} time evolution of the composite system. 
{Given the MGF $G_\lambda(\nu,t)$, the probability distribution $P({\cal E_\lambda},t)$ for the energy transfer amount
${\cal E_\lambda}$ is $ P({\cal E_\lambda},t)  \!=\! \int \! {\rm d}\nu\,G_\lambda(\nu,t) \,{\e}^{-\i \nu  {\cal E_\lambda}} $.
The $n$th derivative of the MGF taken at $\nu=0$ yields the $n$th moment of the energy,
$ \average{{\cal E}_\lambda^{(n)}(t)}= \left.(-\i )^n \d^n   G_\lambda(\nu,  t)/ \d\nu^n \right|_{\nu=0} $. Here we focus on the first moment, i.e., the
energy 
transferred on average, $\average{{\cal E}_\lambda(t)} = \average{{\cal E}_\lambda^{(1)}(t)}$. 
Finally, the control parameter $\lambda$ serves to treat both protocols on equal footing.
For $\lambda =0$, we meet the reservoir measurement $\langle{\cal E}_{\rm R}(t)\rangle$ \cite{gasparinetti2014heat, carrega2015functional}. For $\lambda
=1$, we are actually probing $\langle{\cal E}_{{\rm RI}} (t)\rangle $.}

The expression in Eq. (\ref{eq:gf1}) can be conveniently processed using a functional integral approach \cite{weiss2012quantum, ingold2002path}.
This procedure is a generalization of the one exposed in \cite{carrega2015functional} and the details can be found in the Supplementary Material (SM) \cite{SM}. 
%Here we  focus on the first moment, i.e., the various energies transferred on average. We readily get for $\langle{\cal E}_\lambda(t)\rangle \equiv  \average{{\cal E}_\lambda^{(1)}(t)}$ the path sum representation in terms of the quasi-classical path
{We readily get for $\langle{\cal E}_\lambda(t)\rangle$ the path sum representation in terms of the quasi-classical path}
 $\eta(\tau)=[q(\tau) + q'(\tau)]/q_0$ and  fluctuation path  $\xi(\tau)=[q(\tau)-q'(\tau)]/q_0$,
 where $q_0$ is a  length unit,   as
\beq  \nonumber
 \average{{\cal E}_\lambda ^{} (t)} &=& \int \d \eta_i \,p_{\rm S}^{}(\eta_i)
 \int \d \eta_f     \\    \label{eq:firstmoment}
&&     \times \; \int D \eta \int    \D \xi \, {\e}^{\i S_{\rm S}[\eta,\xi]} \,  {\cal F}_{\rm FV}[\eta,\xi]\, 
\varepsilon^{}_{\lambda}[\eta,\xi]  \, .    \label{eq:average_en}
\eeq
with $\eta_i = \eta(0)$ and $\eta_f = \eta(t)$.
Here ${\cal F}_{\rm FV}[\eta,\xi]$ is the standard Feynman-Vernon influence functional accounting for friction and quantum thermal noise, and  $S_{{\rm S}}[\eta,\xi] = S_{\rm S}[q] -  S_{\rm S}[q'] $ is the system's action related to the double path
$\{\eta,\xi\}$.
The energy functional is found to read
\begin{eqnarray}  \label{eq:Phi1} 
&& \varepsilon^{}_{\lambda}[\eta, \xi] = - \frac{1}{2}\int_0^t \! \d \tau_2 \int_0^{\tau_2 }\!\! \d \tau_1\!\ \dot\eta(\tau_2)   \\
&&\qquad \qquad \quad \times \; \big[ \,  W_1(\tau_2-\tau_1) \dot{\xi}(\tau_1) +\, W_2 (\tau_2-\tau_1)\dot{\eta}(\tau_1) \,\big]   \nonumber  \\    \nonumber
&&\qquad\quad + \int_0^t \! \d \tau_1 \Big\{ -  \dot{\eta}(\tau_1) W_2(\tau_1) \, \eta_i    \\
&& \qquad\quad +(1-\lambda)\eta_f \big[ W_1(t-\tau_1)\dot{\xi}(\tau_1) + W_2(t-\tau_1)\dot{\eta}(\tau_1)  \big]  \Big\}.
\nonumber
\end{eqnarray}
Here $W_1(\tau) =  \dot W'(\tau) $, $ W_2(\tau) = \dot W''(0) - \dot W''(\tau)$, in which 
$W(\tau) = W'(\tau) + \i~  W''(\tau)$ is the  bath correlation function.

The expression (\ref{eq:average_en}) with (\ref{eq:Phi1}) is generally valid for any composite quantum system with bilinear
system-bath coupling. It may serve as a starting point for the implementation of powerful numerical schemes \cite{keil2001realtime,bulla2003numerical,koch2008NonMarkovian,koch2010Semiclassical,orth2013nonperturbative}.

{\it Spin-boson model ---}
To explore  the powerfulness of our approach, consider the dissipative two-state system with two localized states at
a distance $q_0$. In spin representation, the system Hamiltonian reads
\cite{leggett1987dynamics,weiss2012quantum}
\be
\label{eq:sb}
H_{{\rm S}}(t)= {\textstyle  -\frac{1}{2}\Delta\, \sigma_x - \frac{1}{2} \epsilon(t)\,\sigma_z } \, ,
\ee
where $\Delta$ represents the transfer amplitude and $\epsilon(t)$ the time-dependent bias.
The path sum in Eq.~(\ref{eq:firstmoment}) can be written in terms of  all possible sequences of visiting the 
diagonal  states (sojourns) $\eta_j=\pm 1$
and off-diagonal states (blips) $\xi_j=\pm 1$ of the RDM \cite{leggett1987dynamics,weiss2012quantum}. 
% It   takes the form \cite{SM}
\beq    \nonumber
\average{{\cal E}_{{\lambda}} (t)} &=& 
\sum_{m=1}^\infty \Big( \frac{-\Delta^2}{4}\Big)^{m}   \int_0^t  \mathcal{D}_{2 m} \{ t_j\}    \\
& &\times  
\sum_{ \{\xi_j=\pm 1 \}} B_m \sum_{\{\eta_j =\pm 1\}''} {\cal F}_m  \,  \mathfrak{E}_m(\lambda) \, .
\label{eq:Qn1}
\eeq
Here, $ \int_0^t \mathcal{D}_{2 m} \{ t_j\} $ denotes time-ordered integrations of  the $2m$ flip times $t_j$. The sums
$\sum_{ \{\xi_j=\pm 1 \} }$ and $\sum_{\{\eta_j =\pm 1\}'' }$  account for all possible sequences of the $m$ off-diagonal and $m-1$ internal diagonal states of the RDM, respectively. For a given $\{\xi,\eta\}$ configuration of the $2m$ flips, the factor $B_m$ represents the bias  weight, 
and the influence factor ${\cal F}_m$ administers quantum noise and friction (see SM \cite{SM}).
For simplicity, we have implied the initial state $p_{\rm S}^{}(\eta_i) = \delta_{\eta_i,1}$.

The terms governing the energy transfer to the reservoir-plus-interaction,  the reservoir and the  interaction alone are
$\mathfrak{E}_{{\rm RI},m} \equiv\mathfrak{E}_{\lambda=1, m}$, $\mathfrak{E}_{{\rm R},m} \equiv\mathfrak{E}_{\lambda=0, m}$, and
$\mathfrak{E}_{{\rm I},m} =\mathfrak{E}_{{\rm RI}, m}- \mathfrak{E}_{{\rm R},m}$. We readily get  
\beq \nonumber
 \mathfrak{E}_{{\rm I}, m} &=&   \sum_{k=0}^{m-1} \mathcal{U}_{k} (t) \, \eta_k  
+  \i   \sum_{j=1}^{m} \mathcal{V}_{j}(t) \,  \xi_j   \, ,  \\    
 \mathfrak{E}_{{\rm RI}, m} &=& 
 \sum_{k=0}^{m-1}  \mathcal{U}_{k} (t_{2m}) \eta_k  +  \i   \sum_{j=1}^{m} \mathcal{V}_{j} (t_{2m}) \xi_j  \, ,
  \label{eq:etafsum}
  \eeq
where  $\mathcal{U}_{k} (\tau) = \dot{W}''(\tau-t_{2k}) - \dot{W}''(\tau-t_{2k+1}) $, and  where
$\mathcal{V}_{j} (\tau) =\dot{W}'(\tau-t_{2j}) - \dot{W}'(\tau-t_{2j-1})$. 
%In addition, the energy transfer to the reservoir is governed by  $ \mathfrak{E}_{{\rm R}, m}= \mathfrak{E}_{{\rm RI}, m} - \mathfrak{E}_{{\rm I}, m}$.

Interestingly, the expression (\ref{eq:Qn1})  for $\langle{\cal E}_{\rm RI}(t)\rangle$ can be converted, upon integration by parts, into an energy balance relation.
Together with the series expressions analogous to Eq.~(\ref{eq:Qn1}) for the spin functions  $\langle\sigma_x(t)\rangle$  and $\langle\sigma_z(t)\rangle$ \cite{weiss2012quantum}, we get 
\be \label{eq:balance}
\langle {\cal E}_{{\rm RI}}(t) \rangle \equiv \langle {\cal E}_{{\rm R}}(t)\rangle + \langle {\cal E}_{{\rm I}} (t)\rangle = 
 -\langle {\cal E}_{{\rm S}}(t)\rangle + \langle {\cal E}_{{\rm exc}}(t)\rangle\, .
\ee 
Here,  $\langle {\cal E}_{{\rm S}}(t)\rangle= -\frac{1}{2}\Delta \langle \sigma_x(t)\rangle - \frac{1}{2}[\epsilon (t)\langle \sigma_z (t)\rangle - \epsilon (0)]$ is the mean energy transferred to the system, and the contribution $\langle {\cal E}_{{\rm exc}}(t)\rangle= -\frac{1}{2}\int_0^t dt' \dot{\epsilon}(t')\langle \sigma_z (t')\rangle $ is the excess energy pumped into the composite system by the  work spent by the driving.
These results are valid for general linear dissipation.

Next we turn to Ohmic coupling $q_0^2 J(\omega)/\pi =2 K \omega\, \e^{-\omega/\omega_{\rm c}}$.
Here, $K$ is a dimensionless friction strength, and $\omega_{{\rm c}}$  a cut-off. The  Ohmic bath correlation function 
for $\omega_{\rm c}\tau\gg1$ reads~\cite{weiss2012quantum}
\be  \label{eq:strictlyohm}
W(\tau) =    2K\ln\left[ (\beta \omega_{\rm c}/\pi)\sinh( \pi  |\tau|/\beta )\right] 
+ \i ~\,\pi K {\rm sign}(\tau) \, .        
\ee
With this,  $\Delta^2$ is combined with $\omega_{\rm c}$  in the form   $\Delta^2\,\e^{-W'(\tau)}$ as
$\Delta_{\rm r}^{2-2K} =\Delta^2/\omega_{\rm c}^{2K}$. The scaling limit is $\omega_{\rm c}\to \infty$ with
$\Delta_{\rm r}$ fixed.

For general $K$, the series  for the individual  terms of the relation (\ref{eq:balance}) cannot be summed in analytic form. 
However, if $K$ is near $\frac{1}{2}$, say $0.3\lesssim K \lesssim 0.7$, 
the path sums  are feasible using techniques reported in 
Refs.~\cite{sassetti1990, grifoni1993Nonlinear, weiss2012quantum}.
Following these lines,  $\langle\sigma_{x/z}(t)\rangle$ can be written in closed form as
\beq\nonumber
\langle \sigma_z(t)\rangle  &=& F_{\rm S}(t) +   \int_0^t \d\tau \,\,
R(\tau) F_{\rm B}(\tau)\int_0^{t-\tau} \!\! \d s  \\  \nonumber
&&\qquad\qquad \times\,\,F_{\rm S}(s) \sin[\varphi(t-s,t-s-\tau)]    \, ,   \\[1mm]   
\langle \sigma_x(t)\rangle &=& \frac{1}{\Delta} \int_{1/\omega_{\rm c}}^t \! \d\tau\, \,R(\tau) F_{\rm B}(\tau)
\cos[\varphi(t,t-\tau)] \, ,     \label{eq:sigmatoul}
\eeq
where $R(\tau) = \Delta^2\sin(\pi K)\,\e^{-W'(\tau)}$, and $\varphi(t_2,t_1) = \int_{t_1}^{t_2} \d \tau\,\epsilon(\tau)$
is the bias phase accumulated  between $t_1$ and $t_2$. The form factors $F_{\rm B}(\tau)$ and $F_{\rm S}(s)$ 
dress the intervals $\tau$ and $s$, in which the system dwells in an off-diagonal and diagonal state of the RDM, respectively. They obey $F_{\rm B}(0) =F_{\rm S}(0) =1$ and drop to zero on the time scale set by the inverse of the
Kondo frequency, which is  $\Delta_{\rm K} =  [\Gamma(1-K)/2^K]^{1/(1-K)} \Delta_{\rm r}$ for $K$ near $\frac{1}{2}$.

In the same way the series for  $\langle {\cal E}_{{\rm I}}(t)\rangle$ is summed to
\beq\nonumber
 \langle {\cal E}_{{\rm I}}(t)\rangle &=&\frac{1}{2}   \int_0^t \d\tau \,\,
R(\tau) F_{\rm B}(\tau)  \int_{1/\omega_{{\rm c}}}^{t-\tau} \!\!\! \d s\, \, F_{\rm S}(s) \\    \label{eq:Eint}
 &\times &\cos[\varphi(t-s,t-s-\tau)]   [ \dot{W}'(s)-\dot{W}'(\tau+s)] \, .
\eeq
Together with the expressions for $\langle {\cal E}_{{\rm S}}(t)\rangle$ and  $\langle {\cal E}_{{\rm exc}}(t)\rangle$ obtained with (\ref{eq:sigmatoul}),  all constituents of the relation (\ref{eq:balance}) are given.

\begin{figure}
  \begin{center}
    \includegraphics[scale=.7]{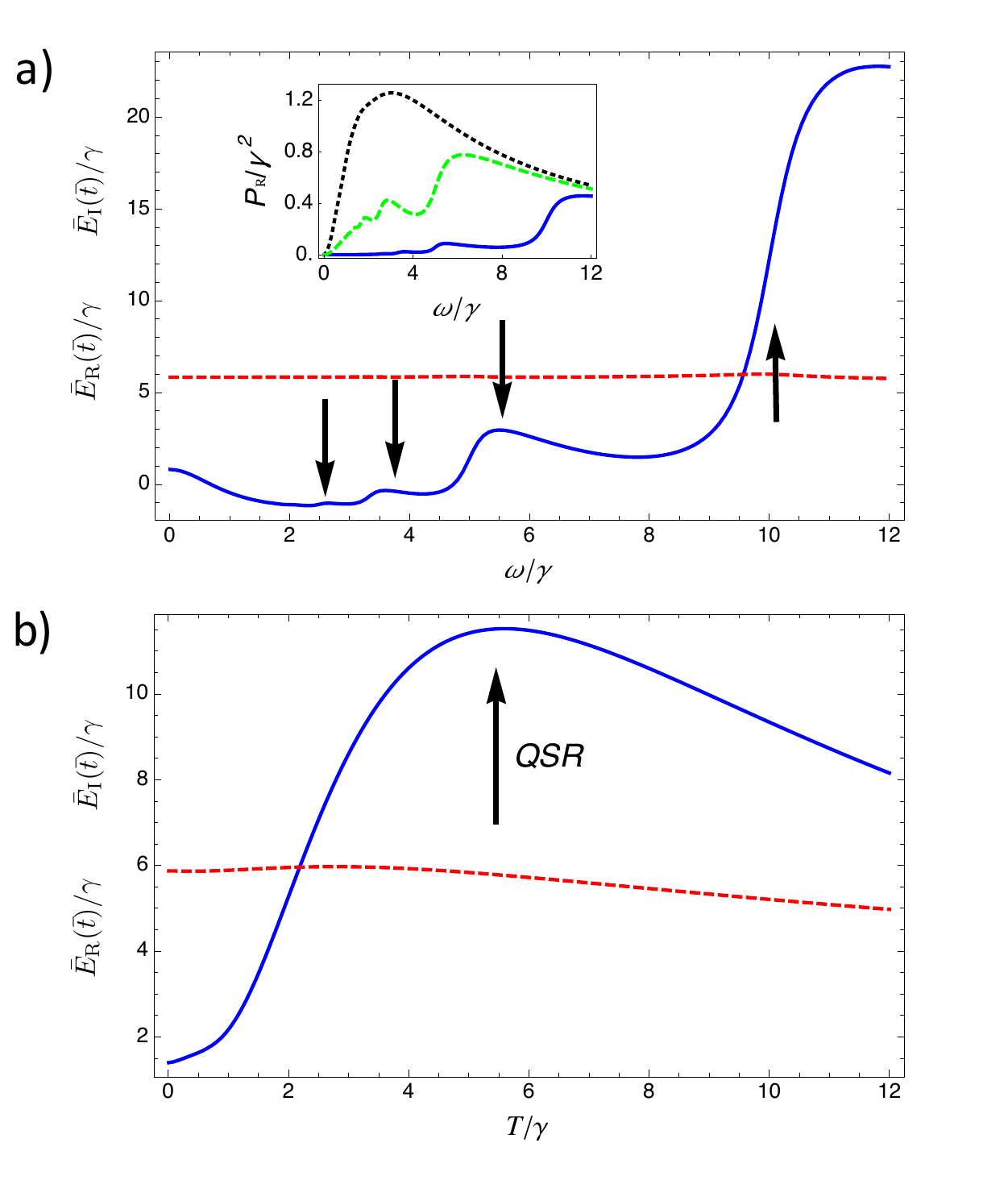}
  \end{center}
 \caption{  (Color online) A snapshot of $\bar{E}_{\rm R} (\bar{t})$ at time $\bar{t}\equiv 50/\gamma$ (solid curve) and 
 $\bar{E}_{\rm I}$ (dashed curve) are plotted versus frequency $\omega$ in a) and versus temperature $T$  in b).  The parameters are
 $\epsilon_0=10 \gamma$, $\epsilon_1=5 \gamma$, $\omega_{\rm c}=5000 \gamma$, plus $T=0.1\gamma$ in a),
 and $\omega =5\gamma$ in b).  In the inset the constant-in-time contribution to the power $P_{\rm R}$ versus $\omega$ is plotted for
 $\epsilon_0=2\epsilon_1$ (solid blue), $\epsilon_0=\epsilon_1$ (dashed green), and $\epsilon_0=0.4 \epsilon_1$ 
 (dotted black). 
The arrows denote the positions of the ground and side frequencies in a) and the quantum stochastic resonance in b).
  }    
    \label{fig:fig1}
\end{figure}

For harmonic driving, $\epsilon(t) =\epsilon_0+ \epsilon_1 \cos(\omega\, t)$, the initially transient dynamics vanishes on the time scale set by the system's relevant relaxation time $\tau_0 = {\rm min}(1/\Delta_{\rm K},\,\beta/\pi)$. In the subsequent stationary regime, the energy transfer contribution $\langle {\cal E}_j(t)\rangle$, where
$ j={\rm S},\,{\rm R},\,{\rm I}, {\rm exc}$,   is found, upon Fourier expansion of the respective bias factor, to read
\be \label{eq:fourseries}
\langle{\cal  E}_j (t)\rangle =  P_{j} \, t   + \sum_{m} E_{j,m} \,\e^{-\i \,m \omega \,t} \, .
\ee
%Explicit forms for the various coefficients may be found with the expressions for $\langle {\cal E}_{{\rm S}}(t)\rangle$, $\langle {\cal E}_{{\rm exc}}(t)\rangle$ and Eq. (\ref{eq:Eint}).
The first term  represents the energy draining into channel $j$ at constant power $P_{j}$ in the period $t$. 
In particular, we obtain 
%\beq \label{eq:P_S}
%P_{\rm \,S} &=& P_{\rm\, I} = 0  \, ,   \\     \nonumber
%P_{\rm R}&=& P_{\rm exc}  =\epsilon_1 \frac{\Delta^2}{2} \frac{\omega^2}{\omega^2+\gamma^2} \int_0^\infty \!\!\! \d\tau \,\,%\e^{-W'(\tau)} F_{\rm B}(\tau)\cos(\epsilon_0\tau) \\
%&&\qquad \;\; \times \;J_1[z(\tau)]\left[ \cos\left(\frac{\omega\tau}{2}\right)+ \frac{\gamma}{\omega}
% \sin\left(\frac{\omega\tau}{2}\right)\right]    \, ,
 %\label{eq:P_R}
%\eeq
\beq \nonumber
P_{\rm \,S} &=& P_{\rm\, I} = 0  \, ,   \\     \nonumber
P_{\rm R}&=& P_{\rm exc}  =\epsilon_1 \frac{\omega}{2} \int_0^\infty \d\tau \,R(\tau) F_{\rm B}(\tau)\cos(\epsilon_0\tau)
J_1[z(\tau)]  \\
&&\qquad \quad\times \;  \int_0^\infty \d s\, F_{\rm S}(s) \sin\left[\omega\left(\frac{\tau}{2}+s\right) \right]   \, ,
 \label{eq:P_R}
\eeq
where $z(\tau)= 2 \epsilon_1\sin(\omega\tau/2)/\omega$, and  $J_1(x)$ is a Bessel function.
The term $P_{\rm exc}$ is the constant part of the power injected into the composite system by the drive in the stationary state. Evidently, this amount is fully absorbed by  the reservoir.

With the results (\ref{eq:P_R})
one directly sees from Eq.~(\ref{eq:fourseries}), that in the long run
the reservoir contribution $\langle{\cal  E}_{\rm R} (t)\rangle$ dominates over  the interaction one, and lastly the energy is predominantly dissipated in the environment.
Nevertheless, there are intermediate stationary time regimes in which the reservoir and interaction contributions are of the same order of magnitude.

Let us now look more closely at the particular case $K=\frac{1}{2}$, in which the form factors can be calculated exactly in analytic form, yielding
$F_{\rm B}(\tau) = \e^{-\gamma\tau/2}$ and $F_{\rm S}(s) = \e^{-\gamma s}$ with the Kondo frequency
$\gamma \equiv \Delta_{\rm K}(K=\frac{1}{2}) = \frac{1}{2} \pi\Delta^2/\omega_{\rm c}$~\cite{sassetti1990, weiss2012quantum}.

 Consider the energy flow until time $t$. Taking in (\ref{eq:fourseries})  the average over period 
 $2\pi/\omega$,  we get with (\ref{eq:balance}) $\bar{E}_{{\rm I}}(t) = E_{{\rm I},0}$ and
$\bar{E}_{\rm R}(t) =  P_{\rm \,R}\, t - E_{{\rm I},0}  - E_{{\rm S},0} +  E_{{\rm exc},0}$. In practice, this corresponds to
 the case, in which the oscillating terms are averaged out during the measurement, i.e., by using a detector
  unable to resolve energy variation within  period $2\pi/\omega$.
In the scaling regime, the leading asymptotic behaviors are 
$E_{{\rm I},0} = \frac{\gamma}{2\pi } [ \log(\omega_{\rm c}/\gamma)]^2 $ and 
$E_{{\rm S},0}  = -\frac{\gamma}{2}\ln(\omega_{\rm c}/\gamma)$, while $E_{{\rm exc},0}$ does not depend on 
$\omega_{\rm c}$. Hence we have in the scaling regime $E_{{\rm I},0}  \gg  |E_{{\rm S},0}]\gg E_{{\rm exc},0} $,  and thus
$\bar{E}_{\rm R}(t) \simeq  P_{\rm \,R}\, t - E_{{\rm I},0}$.

To understand how the energy is dissipated into the various channels, we plot in Fig.~\ref{fig:fig1} 
a snapshot of $\bar{E}_{\rm R}(\bar{t})$ and $\bar{E}_{{\rm I}}(\bar{t})$ at time 
 $\bar{t}\equiv 50/\gamma$ as functions of frequency $\omega$ and of temperature $T$.
 The behavior of the reservoir channel is particularly interesting as it shows a sequence of plateaus combined with
 sudden ascents around $\omega=\epsilon_0/n$, $n=1,\,2,\,3,\,4$.
 This resembles the driving-induced coherences (DIC) and resonances inherent in strongly coupled system-reservoir entities \cite{grifoni1996coherent,  grifoni1998driven, kast2013persistence, kast2013dynamics}.
The resonances  fade away into the continuum when  $\epsilon_0 < \epsilon_1$ is small [see inset in 
Fig.~\ref{fig:fig1} a)], or when temperature is increased.

The behavior of  $\bar{E}_{\rm R}(\bar{t})$  reflects the one of the power $P_{{\rm R}}$, as depicted in the inset of Fig.~\ref{fig:fig1} a).
The solid curve, reporting $ \bar{E}_{\rm R}(\bar{t})$, is effectively an amplification of the solid curve in the inset,
showing $P_{\rm R}$,  by a factor $50/\gamma$ with a displacement along the ordinate by roughly $- E_{{\rm I},0} $. 
We see that in a region around $\omega=2\gamma$  there holds $\bar{E}_{\rm R}(\bar{t}) <0$, i.e.,
  the energy $E_{{\rm I},0} $ drained away into the interaction still outstrips the amount $P_{\rm R}\bar{t}$. 
  Since $\bar{E}_{\rm R}(\bar{t})$ depends strongly on  $\omega$ and $\epsilon_0$, and $\bar{E}_{{\rm I}}(t) $ is virtually constant, the case, in which the energy dissipated into the bath is small or large,  can easily be adjusted by tuning these external parameters.
  %The inset of Fig.~\ref{fig:fig1}  shows that the power $ P_{\rm R}(\omega)$ is increasingly reduced for all $\omega$, as the static bias is tuned from $\epsilon_0\ll\epsilon_1$ to  $\epsilon_0\gg \epsilon_1$. 

\begin{figure}
    \begin{center}
    \includegraphics[scale=.7]{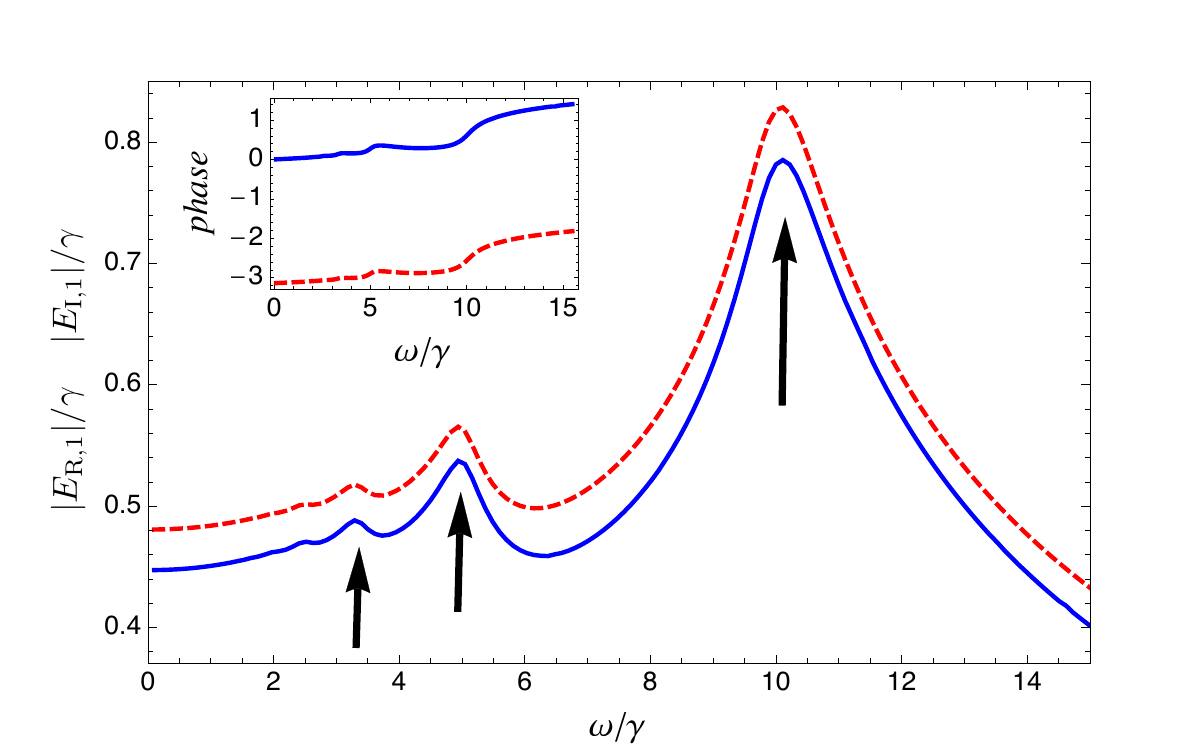}
      \end{center}
    \caption{        
     (Color online) The absolute values of the amplitudes of the first harmonic $E_{\rm \, R,1}$ (solid curve) and 
     $E_{\rm \,I,1}$ (dashed curve) are plotted versus driving frequency $\omega$. The parameters are the same as in 
     Fig.~\ref{fig:fig1} a). Both curves show resonances at $\epsilon_0/n$ (indicated by the arrows).
    Inset: phases of $E_{\rm \, R,1}$ (solid curve) and $E_{\rm \,I,1}$ (dashed curve) versus $\omega$.
    The two contributions are always in anti-phase. }
    \label{fig:fig2}
\end{figure}

The energies $\bar{E}_{\rm \, R}(\bar{t})$ and $\bar{E}_{\rm I}(\bar{t})$ are depicted versus temperature 
in Fig.~\ref{fig:fig1} b). While the latter depends weakly on  $T$, the energy deposited in the bath has a peaked structure signifying that the respective energy is amplified because of the noise. This reflects the quantum stochastic resonance characteristics of the driven damped quantum system \cite{grifoni1996coherent}. 
Side resonances are absent because of the relatively high temperatures. At low $T$, the curve for $\bar{E}_{\rm\, R}(\bar{t})$ falls again below $\bar{E}_{{\rm I}}(\bar{t})$.

Finally, we turn to the oscillatory part in Eq.~(\ref{eq:fourseries}). The individual terms of the $m$th component represent
oscillatory energy exchange between the four entries of the balance relation (\ref{eq:balance}) at frequency $m\omega$.
These features can be observed with a  detector having time resolution smaller than $2\pi/\omega$.
The amplitudes of the  $m=\pm 1$ components are the largest ones.
The absolute values of the amplitudes $E_{\rm \,R,1}$ and $E_{\rm \, I,1}$  are depicted versus $\omega$ for a
particular parameter set in Fig.~\ref{fig:fig2}. The amplitudes  show distinct resonances at $\omega =\epsilon_0/n$ with
$n=1,2$, and 3 which are again related to the DIC.
Since the phases of the amplitudes differ by nearly $\pi$, as shown in the inset of Fig.~\ref{fig:fig2}, 
and the absolute values are close, the major energy transfer is between the bath and the interaction.
The surplus of energy put in and out of the reservoir oscillates with amplitude 
$E_{\rm exc,1}-E_{\rm S,1}$, and the absolute value of it is just the difference of the two curves in Fig.~\ref{fig:fig2}.
This is small since the leading contributions to $E_{\rm exc,1}$ and $E_{\rm S,1}$ cancel each other.

For completeness, consider the asymptotic behaviors of the various energy contributions in the scaling regime 
for $K\neq \frac{1}{2}$.
For $K<\frac{1}{2}$, we obtain from Eq.~(\ref{eq:Eint}) $E_{{\rm I},0}\propto \ln(\omega_{\rm c}/\Delta_{\rm r})$, while $E_{{\rm S,0}}$ and $E_{{\rm exc,0}}$ are independent of $\omega_{\rm c}$. 
For $K>\frac{1}{2}$, we get  from Eq.~(\ref{eq:sigmatoul}) 
 $E_{{\rm S},0} = - c_{{\rm S},0}^{}\, \Delta_{\rm r} (\omega_{\rm c}/\Delta_{\rm r})^{2K-1}$,  and from 
 Eq.~(\ref{eq:Eint}) the same functional form, $E_{{\rm I},0} = c_{{\rm I},0}^{} \,\Delta_{\rm r} (\omega_{\rm c}/\Delta_{\rm r})^{2K-1}$ with the ratio
$c_{{\rm I,0}}/c_{{\rm S},0}= 2\pi K/\sin[2\pi(K-\frac{1}{2})]$.  Interestingly, this term is very large for $K$ slightly 
above $\frac{1}{2}$. Thus for $K$ around $\frac{1}{2}$, we generally have  
$E_{{\rm I},0}\gg |E_{{\rm S},0}|,\, E_{{\rm exc},0}$. Hence the characteristic behaviors depicted in Fig.~\ref{fig:fig1}
qualitatively hold for $K$ near $\frac{1}{2}$. However, the time at which the energy drained into the bath exceeds the energy 
$E_{{\rm I},0}$ drained into the interaction may sensitively depend on whether $K< \frac{1}{2}$ or $K>\frac{1}{2}$. 
Finally, since all the oscillatory terms
in Eq.~(\ref{eq:fourseries}) are independent of $\omega_{\rm c}$, the coefficients $E_{j,m}$, where $j={\rm S, \, R,\,I}$, 
and ${\rm exc}$,
smoothly vary with $K$ near $K=\frac{1}{2}$.

{\it Conclusions ---}
We presented a general method to analyze the time-resolved energy transfer to the various dissipative channels  of a driven open quantum system strongly coupled to a heat bath. The exact formal solution in path sum representation was given in 
Eqs.~(\ref{eq:average_en}) and (\ref{eq:Phi1}). It may form a firm basis for the implementation of efficient numerical 
tools \cite{keil2001realtime,bulla2003numerical,koch2008NonMarkovian,koch2010Semiclassical,orth2013nonperturbative}.
The method was applied to the spin-boson model, with explicit results in the Ohmic scaling limit at strong coupling, for the special case of $K=\frac{1}{2}$.
We showed that the interaction channel is a relevant dissipative drain which even may dominate in particular regimes of the parameters. For harmonic driving, the energy flowing into the reservoir shows distinct resonant behavior, thereby reflecting quantum stochastic resonance features of the model \cite{grifoni1996coherent,grifoni1998driven}.
The findings may open new directions in the study of energy transfer in complex quantum systems, and pave the way
to control energy dissipation into the reservoir by tuning the system's parameters.

\begin{acknowledgments}
M.C., P.S. and M.S. acknowledge the support of the MIUR-FIRB2012 - Project HybridNanoDev (Grant  No.RBFR1236VV), EU FP7/2007-2013 under REA grant agreement no 630925 -- COHEAT, MIUR-FIRB2013 -- Project Coca (Grant No.~RBFR1379UX), and the COST Action MP1209.
U. W. acknowledges support from the Deutsche Forschungsgemeinschaft through SFB/TRR21.
\end{acknowledgments}

%%%%%%%%%%%%%%%%%%%%%%%%%%%%%%%%%%%%%%%%%%%%%%%%%%%%%%%%%%%%
%\bibliographystyle{apsrev4-1}
%\bibliography{strong_coupling_biblio}
%%%%%%%%%%%%%%%%%%%%%%%%%%%%%%%%%%%%%%%%%%%%%%%%%%%%%%%%%%%%
%

%%%%%%%%%%%%%%%%%%%%%%%%%%%%%%%%%%%%%%%%%%%%%%%%%%%%%%%%%%%%
% Supplementary material
%%%%%%%%%%%%%%%%%%%%%%%%%%%%%%%%%%%%%%%%%%%%%%%%%%%%%%%%%%%%

\appendix
\newpage
\setcounter{equation}{0}

\section{Supplemental Material}

Initially, we outline the derivation of the moment generating function (MGF) $G_\lambda (\nu,t)$ within the functional integral approach. 
After that we give some details relevant for the application to the spin-boson model.

The total Hamiltonian of the  system-plus-reservoir is $H(t) = H_{\rm S}(t)+ H_{\rm R}+ H_{\rm I}$~\cite{caldeira1983path, weiss2012quantum, ingold2002path, leggett1987dynamics}.
We assume that at times $t\leq 0$, the system and the bath are decoupled, the system dwells at position $q(0) \equiv q_i$
with probability $p_{{\rm S}}(q_i)$, and the bath is in thermal equilibrium  at inverse temperature $\beta=1/T$ (hereafter we set $\hbar=k_{{\rm B}}=1$). 
Hence we have
$\langle q_i | \rho(0)|q_i \rangle = p_{{\rm S}}(q_i) \e^{- \beta H_{{\rm R}}}/Z_{{\rm R}}$ with 
$Z_{{\rm R}}={\rm Tr} \,\e^{-\beta H_{{\rm R}}} $.
We are interested in  the MGF, which includes the entire
statistics of the energy exchange process according to the measurement protocols discussed in the Letter.
Formally the MGF can be written as a generalized density matrix \cite{gasparinetti2014heat}
\begin{equation} \label{eq:gf1}
  G_\lambda (\nu,t) = \Trace\, [ \e^{\i \nu( H_{\rm R}+\lambda H_{\rm I})} U_{t, 0} \,
\e^{- \i  \nu  H_{\rm R} }\rho(0) U^\dagger _{t, 0} ] \, ,
\end{equation}
where $U_{t,0}$ conveys the time evolution of the total system.

It is expedient to perform the reduction of the dynamics of the composite system to that of the system alone  
using the  path sum approach \cite{weiss2012quantum, carrega2015functional}. Within this framework, we have
\beq
\label{eq:sm1}
G_\lambda (\nu, t) &=& \int \d q_i\,  p_{{\rm S}}^{}(q_i) \int \d q_f 
\int\limits_{q(0)=q_i}^{q(t)=q_f} \D q    \int\limits_{q'(0)=q_i}^{q'(t)=q_f} \D q' 
\nonumber \\    
&&\times \; \e^{\i (S_{{\rm S}}[q] - S_{{\rm S}}[q'])} {\cal F}_\lambda [q,q';\nu]~,
\eeq
{where $S_{{\rm S}}[q]$ represents the action of the system for the path $q(\tau)$.}
The influence functional ${\cal F}_\lambda [q,q';\nu]$ is  given by
\beq   \nonumber
&&\mathcal{F}_\lambda [q,q';\nu]=Z_{\rm R}^{-1}\int \d {\boldsymbol x} \,
\matrixel{{\boldsymbol x}'_f}{\e^{\i \nu( H_{\rm R}+\lambda H_{\rm I})     
}_{} }{{\boldsymbol x}_f}\,      \\  \label{eq:sm2}
&&\qquad \quad \times\; F^*[q', {\boldsymbol x}'_f, {\boldsymbol x}'_i] \, 
F[q,{\boldsymbol x}_f, {\boldsymbol x}_i]   \,
 \matrixel{{\boldsymbol x}_i}{ \e^{- ( \beta+ \i \nu) H_{\rm R} }_{}    }{{\boldsymbol x}'_i} \, .
 \eeq
Here, the vector ${\boldsymbol x}'_i$ combines  the reservoir coordinates $(x'_{1,i}, ....,x'_{N,i})$, 
$ \int  \d {\boldsymbol x} = \int \d {\boldsymbol x}_f \d {\boldsymbol x}'_f  \d {\boldsymbol x}_i \d {\boldsymbol x}'_i $,
and $q$ is the position of the system.
We have introduced the amplitude
$F[q,{\boldsymbol x}_f, {\boldsymbol x}_i] =\int_{
{\boldsymbol x}_i}^{{\boldsymbol x}_f} \!\! \D{\boldsymbol  x} \,
{\e}^{\i  ( S_{\rm R}[{\boldsymbol x}]+S_{\rm I}[q,{\boldsymbol x}]  )} $, 
in which $S_{\rm I}[q,{\boldsymbol x}] = - \int d\tau\,H_{\rm I}$
.
The  influence functional ${\cal F}_\lambda [q, q'; \nu]$ accounts for all environmental effects on the energy transfer statistics captured by the MGF $G_\lambda (\nu,t)$. 
Since the integrand of (\ref{eq:sm2}) is a mixed quadratic form of all the bath variables, the multiple integrations 
can be done in closed form upon completing the squares.

It is convenient to rewrite Eq.~(\ref{eq:sm2}) in terms of quasi-classical  path $\eta(\tau )=(q(\tau)+q'(\tau))/q_0$ and quantum fluctuation path $\xi(\tau)=(q(\tau)  - q'(\tau))/q_0$, with $q_0$ being  a characteristic length of the system.
Then the MGF takes the form
\beq  \nonumber
 G_\lambda (\nu,t) &=& \int \d \eta_i  p_{{\rm S}}(\eta_i)\int \d \eta_f  
 \int\limits_{\eta(0)=\eta_i}^{\eta(t)=\eta_f}  \!\!  \!\!  \D\eta \int\limits_{\xi(0)=0}^{\xi(t)=0} \D\xi    \\
&&\qquad \times \; \e^{ \i S_{{\rm S}}[\eta,\xi]} {\cal F}_\lambda [\eta , \xi ;\nu]
\eeq
with the influence functional
\be
  {\cal F}_\lambda [\eta,\xi;\nu] = {\cal F}_{\rm FV}[\eta,\xi]\;  \e^{ \i \Phi_{\lambda}[\eta,\xi;\nu] } \, .
\ee
Here ${\cal F}_{\rm FV}[\eta,\xi]$ is the  Feynman-Vernon influence functional accounting for quantum noise and friction \cite{weiss2012quantum, carrega2015functional}.  The influence action  
$\Phi_{\lambda}[\eta,\xi;\nu] $ controls the full  statistics of the energy transfer process. It is found to read
\beq \nonumber
&& \Phi_{\lambda }[\eta, \xi;\nu]= \frac{1}{2}  \int_0^t \! \d t_2\!\!\int_0^{t_2 }\!\! \d t_1\!  \\  \nonumber
&&\qquad \times \; \left\{ \big[ - \dot{\eta}(t_2) \dot{\eta}(t_1) + \dot{\xi}(t_2) \dot{\xi}(t_1)\big] W_2 (t_2-t_1,\nu)\right.     \\    \nonumber
&&  \qquad \qquad\left. +\;  \big[- \dot{\eta}(t_2)\dot{\xi}(t_1) +\dot{\xi}(t_2) \dot{\eta}(t_1) \big] 
W_1 (t_2-t_1,\nu)\right\}   \nonumber  \\    \nonumber
&& + \;(1-\lambda)
\eta_f \int_0^t \! \d t_1 \,\big[  W_2(t-t_1,\nu)\dot{\eta}(t_1) +
  W_1(t-t_1,\nu)\dot{\xi}(t_1)   \big]    \\   
&& \qquad+\;  \int_0^t \! \d t_1 \,\big[ - \dot{\eta}(t_1) W_2(t_1,\nu) +
 \dot{\xi}(t_1) W_1(t_1,\nu)  \big] \,  \eta_i  \,   .      \label{eq:PhiQ} 
\eeq
In the above equation we have introduced the kernels
\beq  \nonumber
W_1 (\tau,\nu)\! &=&\! \frac{q_0^2}{\pi} \int \!\! \d \omega
\frac{J(\omega)}{\omega^2}  \frac{f_1(\omega,\nu)}{\sinh(\beta \omega/2) }       \sin(\omega \tau)   \, ,   
    \\
 W_2 (\tau,\nu) \!&=& \! \frac{q_0^2}{\pi} \int \! \! \d \omega
\frac{J(\omega)}{\omega^2} \frac{ f_2(\omega,\nu)}{\sinh(\beta \omega/2) }  [1- \cos(\omega \tau)] \,  ,
           \label{eq:Lnu12} 
\eeq
with 
\beq  \nonumber
f_1(\omega,\nu) &=& 2 \sin(\nu\omega/2)\cosh[(\beta +\i \nu)\omega/2] \,  ,   \\
f_2(\omega,\nu) &=& 2  \sin(\nu\omega/2)\sinh[(\beta +\i \nu)\omega/2]   \, .   \label{eq:ffunc}
\eeq
The  $n$th moment  $\average{{\cal E}_\lambda^{(n)} (t)}$  of the probability distribution
$P({\cal E}_\lambda,t)= \int \d\nu \,G_\lambda(\nu,t)\,\e^{-\i \nu{\cal E}_\lambda}$ is readily found  with the relation
$\langle{\cal E}_\lambda ^{(n)} (t)\rangle = (-\i)^n \d^nG_\lambda(\nu,t)/\d\nu^n|_{\nu=0}$ as
\beq  \nonumber
 \average{{\cal E}_\lambda^{(n)} (t)} &=& \int \d \eta_i \,p_{\rm S}^{}(\eta_i)
 \int \d \eta_f   \!\!\!
\int\limits_{\eta(0)=\eta_i}^{\eta(t)=\eta_f}  \!\!  \D \eta \int\limits_{\xi(0)=0}^{\xi(t)=0}
\!\!    \D \xi\;   \\    
&&\qquad \times \; {\e}^{\i S_{\rm S}[\eta,\xi]} \; {\cal F}_{\rm FV}[\eta,\xi]\,
\varepsilon_\lambda ^{(n)}[\eta,\xi]   \, ,      \label{eq:nthmoment}
\eeq
with
${\varepsilon_\lambda ^{(n)}}[\eta,\xi] =(-\i)^n\, \left. \d^n\, \e^{\i \Phi_{\lambda}[\eta,\xi;\nu]} \d\nu^n \right|_{\nu=0} $.
For $n=0$,  in which $\varepsilon_\lambda ^{(0)}[\eta,\xi]=1$, Eq.~(\ref{eq:nthmoment}) reduces to the normalization condition of the RDM, and thus $\langle {\cal E}_\lambda ^{(0)}(t)\rangle=1$.
This shows that only  terms in the functional $\varepsilon_\lambda ^{(n)}[\eta,\xi]$  which depend on $\eta_f$ 
actually contribute to  $\average{{\cal E}_\lambda ^{(n)} (t)}$.
The kernels entering the functional for the mean energy $\langle {\cal E}_\lambda (t) \rangle \equiv \average{{\cal E}_\lambda ^{(1)}(t)}$ are
$W_{1,2}(\tau) = \partial W_{1,2}(\tau,\nu)/\partial\nu\,|_{\nu=0}$, yielding $W_1(\tau) = \dot{W}'(\tau)$ and $W_2(\tau) =  \dot{W}''(0)-  \dot{W}''(\tau)$, where
$W(\tau)= W'(\tau) + i\,W''(\tau)$ is the bath correlation function
\be  \label{eq:bathcorr}
W(\tau) \!\!\! = \frac{q_0^2}{\pi} \!\!\int \! \d\omega\frac{J(\omega)}{\omega^2} 
\frac{\cosh[\omega\frac{\beta}{2}]-\cosh[\omega(\frac{\beta}{2}-\i \tau)]}{\sinh[\omega\frac{\beta}{2}]}\,.
\ee
The resulting expression for $\langle {\cal E}_\lambda (t) \rangle$ is given in Eq.~(2) of the Letter.

Next, we specify the bias factor $B_m$ and the influence factor ${\cal F}_m$, which enter the path sum expression for the mean energy $\langle {\cal E}_\lambda (t) \rangle$
of the spin-boson model given in Eq.~(5) of the Letter.
The system is sensitive to the bias during  time intervals in which it dwells in an off-diagonal state of the RDM. Labeling the off-diagonal states by $\xi_j=\pm 1$, the bias factor
reads
\be \label{eq:bias}
B_m = \exp\{ \i \,\sum_{k=1}^m \xi_k \varphi(t_{2k},t_{2k-1}) \} \, .
\ee
Here,  $\varphi(t_2,t_1)$ is the phase picked up between times $t_1$ and $t_2$.

The influence function  ${\cal F}_m = G_m H_m$ carries quantum noise and friction in the factors $G_m$ and $H_m$, respectively. Putting $W_{j,k}= W(t_j-t_k)$, we have
\beq \nonumber
G_m &=& \exp\Big(- \sum_{j=1}^m W'_{2j,2j-1}\Big) \;\exp\Big( - \sum_{j=2}^m \sum_{k=1}^{j-1} \xi_j \xi_k \Lambda_{j,k} \Big) \, ,  \\ 
H_m &=& \exp\Big( \i \! \sum_{k=1}^{m-1}\sum_{j=k+1}^{m}  \! \xi_j  X_{j,k}\, \eta_k\Big)   \, .  \label{eq:noisefric}
\eeq
The first  factor in $G_m$ represents the intra-blip correlations, and the second factor the intra-blip correlations, in which
$\Lambda_{j,k} = W'_{2j,2 k-1} +W'_{2 j-1,2k}- W'_{2j,2k}-W'_{2j-1,2k-1}$ is the correlation of blip pair $\{ j,k \}$. 
Finally, the factor $H_m$ describes the correlations of the quasi-classical path with the fluctuation path, in which
$X_{j,k} =W''_{2j,2k+1}+W''_{2j-1,2k}-W''_{2j,2k}-W''_{2j-1,2k+1}$ is the correlation of the blip-sojourn pair $\{ j,k \}$.

%%%%%%%%%%%%%%%%%%%%%%%%%%%%%%%%%%%%%%%%%%%%%%%%%%%%%%%%%%%%
%\bibliographystyle{apsrev4-1}
%\bibliography{strong_coupling_biblio}
%%%%%%%%%%%%%%%%%%%%%%%%%%%%%%%%%%%%%%%%%%%%%%%%%%%%%%%%%%%%
%

%%%%%%%%%%%%%%%%%%%%%%%%%%%%%%%%%%%%%%%%%%%%%%%%%%%%%%%%%%%%

%%%%%%%%%%%%%%%%%%%%%%%%%%%%%%%%%%%%%%%%%%%%%%%%%%%%%%%%%%%%
\end{document}